# Time-varying tip-sample force measurements confirm steady-state dynamics in tapping-mode atomic force microscopy

Ozgur Sahin[a]

The Rowland Institute at Harvard, Harvard University, Cambridge, MA 02142 USA

Direct time-varying tip-sample force measurements by torsional harmonic cantilevers facilitate detailed investigations of the cantilever dynamics in tapping-mode atomic force microscopy. Here we report experimental evidence that the mathematical relationships describing the steady state dynamics are quantitatively satisfied by the independent measurements of tip-sample forces over a broad range of experimental conditions. These results confirm the existing understanding of the tapping-mode atomic force microscopy and build confidence on the reliability of time-varying tip-sample force measurements by the torsional harmonic cantilevers.

PACS. 07.79.Lh, 68.37.Ps,

[a] Corresponding author: sahin@rowland.harvard.edu

# I. INTRODUCTION

Tapping-mode is the most common imaging modality used in atomic force microscopes (AFM) [1]. In this mode, the force sensing cantilever is vibrated at or near its resonance frequency near the vicinity of the sample while the sharp tip forms intermittent contacts with the surface. Relatively large vibration amplitudes prevent sticking, and intermittent contacts minimize damage to the sample and the tip. Besides its practical advantages in imaging, due to the dynamic tip-sample interaction, the tapping-mode offers great potential to characterize and map material properties of samples with high spatial resolution [2].

The advantage of the tapping-mode in characterizing material properties comes with complicated cantilever dynamics due to the non-linear tip-sample interaction [3,4]. The vibrating cantilever exhibits multiple steady oscillation states [5-7], higher-harmonic generation [8], sub-harmonic generation, and chaotic behavior [9,10]. A wide range of experimental results can be described by modeling the cantilever with a damped harmonic oscillator interacting with a non-linear tip-sample potential. Equations describing the steady state dynamics derived from this model predict the energy dissipation during tip-sample interaction as well as magnitudes of average interaction forces [11-13]. Furthermore, the use of amplitude and phase response has been proposed to recover tip-sample interaction potentials [14,15] and identify nanoscale dissipation processes [16]. Beyond the use of the vibration amplitude and phase at the drive frequency, response of higher order modes and higher-harmonic vibrations have been studied to provide additional image contrast mechanisms based on material properties [17-23].

Recently, torsional harmonic cantilevers (THC) are introduced to measure time-varying tip-sample forces in tapping-mode atomic force microscopy [24]. These cantilevers have their tips at an offset distance from the longitudinal axis of the cantilever, so that tip-sample forces excite torsional vibrations on the cantilever. The sensitivity and bandwidth of the torsional vibrations enable measurements of the attractive and repulsive forces and their variation with time or tip-sample separation. In this work we investigate whether the independent tip-sample forces measured by the THCs quantitatively satisfy the steady state equations over a range of tapping conditions. We summarize the steady state equations for the tapping cantilever and discuss the basis for the experimental comparison in the theory section. Then, we present time-varying tip-sample forces under various experimental conditions and show that these forces quantitatively satisfy the steady-state equations. We also discuss the behavior of the measured tip-sample force waveforms in attractive and repulsive regimes and how they agree to the established understanding of the tapping mode.

## II. THEORY

Several formulations of the steady state dynamics of the tapping cantilever have been reported. We will be using the approach that assumes the total harmonic force (driving force plus the first harmonic of the tip-sample force), the vibration amplitude, and phase have to satisfy the relationships for a damped harmonic oscillator [25]. Holscher *et al.* used this approach and provided two coupled equations describing the cantilever motion [26]. Similarly, Sahin et al., provided one equation between complex variables [19]. We will use the equation between complex variables to allow visualization in polar coordinates with phasors. In this form the equation describing the total force acting on the cantilever is as follows:

$$F_T e^{i(\omega t + \beta(\omega))} = F_D e^{i(\omega t + \phi)} + F_{ts1} e^{i(\omega t + \theta)} \qquad (1)$$

Here $F_T$, $F_D$, and $F_{ts1}$ are the magnitudes of the total harmonic force, driving force, and the first harmonic of the tip-sample force, respectively. By modeling the cantilever as a damped harmonic oscillator $F_T$ can be written in terms of experimentally accessible parameters as follows:

$$F_T(\omega) = K_1 A_s \left\{ 1 - \left(2 - \frac{1}{Q^2}\right)\left(\frac{\omega}{\omega_0}\right)^2 + \left(\frac{\omega}{\omega_0}\right)^4 \right\}^{1/2} = K_1 A_s T(\omega), \tag{2}$$

Here $\omega$ is drive frequency and $\omega_0$ is the resonance frequency, $Q$ is the quality factor of fundamental resonance, $K_1$ is the effective spring constant of the fundamental flexural mode, $A_s$ is the vibration (set-point) amplitude. The transfer function term $T(\omega)$ is introduced to simplify the representation of the frequency dependent terms in this and subsequent equations. The corresponding phase $\beta(\omega)$ in Eq. (1) is also determined by the damped harmonic oscillator response and it is given with the following equation:

$$\tan(\beta(\omega)) = \frac{\omega \omega_0 / Q}{\omega^2 - \omega_0^2}, \quad \beta \in \{0, \pi\}. \tag{3}$$

$F_D$ can be written in terms of the free vibration amplitude $A_0$ as follows:

$$F_D(\omega) = K_1 A_0 \left\{ 1 - \left(2 - \frac{1}{Q^2}\right)\left(\frac{\omega}{\omega_0}\right)^2 + \left(\frac{\omega}{\omega_0}\right)^4 \right\}^{1/2} = K_1 A_0 T(\omega), \tag{4}$$

The corresponding phase value for the driving force $\phi$ is the phase difference between the driving force and the cantilever motion (reference).

The third term in Eq. (1) represents the first harmonic of the tip-sample interaction forces. It is given by the Fourier integral of the time-varying tip-sample force $f_{ts}(t)$ as follows;

$$F_{ts1} e^{i\theta} = \frac{\omega}{2\pi} \int_{2\pi/\omega} f_{ts}(t) e^{-i\omega t} dt \tag{5}$$

Upon substitution of equations (2-5) into Eq. (1) we get the relationship that determines $F_{ts1}$ and $\theta$:

$$K_1 A_S T(\omega) e^{i\beta(\omega)} = K_1 A_0 T(\omega) e^{i\phi} + F_{ts1} e^{i\theta}. \quad (6)$$

This equation relates the cantilever vibration to the tip-sample forces in the steady state. We will represent the terms in Eq. (6) with phasors on the complex plane and compare the total force $\mathbf{F_T}$ (first term) to the vector sum of the driving force $\mathbf{F_D}$ (second term) and tip-sample interaction force $\mathbf{F_{ts1}}$ (third term). This representation gives a visual picture on how Eq. (6) is satisfied, but more importantly it shows the relative contribution of attractive and repulsive forces as well as dissipative interactions in how accurately the equation is satisfied. This is especially important as we are going to test Eq. (6) under a range of experimental conditions that includes attractive and repulsive interaction regimes, different set-point amplitudes, and drive frequencies.

## III. RESULTS AND DISCUSSION

Independent measurements of tip-sample interaction forces are performed with a THC that has a fundamental resonance frequency of 70.16 KHz with near surface quality factor equal to 45 and a torsional resonance frequency of 1050.0 KHz with a torsional mode quality factor of 1100. The nominal width and length of the cantilever are 30 and 275 um, respectively. The tip offset distance is 25 um. The spring constant of the fundamental mode is 4.4 N/m, calibrated against thermo-mechanical motion. Time-varying tip sample forces are obtained by recording the lateral bending signal with a 12 bit data acquisition card (National Instruments S-6115) and processing the signal to correct for the distortions due to the torsional resonance and cross-talk from large vertical signal according to the procedures described in ref [24]. Tip-sample force waveform measurements are averaged over 120 oscillation cycles. This gives a measurement bandwidth of 600 Hz. Calibration of the torsional deflection signals (volts to Newtons conversion) is performed according to ref [27]. After obtaining a calibrated tip-sample force waveform, $F_{ts1}$ and $\theta$ are obtained by Eq. (5).

In order to test Eq. (6) over a wide range of conditions, we have recorded amplitude vs. distance and phase vs. distance curves with the cantilever described above at three drive frequencies $\omega_{low}$ = 69.48 KHz, $\omega_{res}$ = 70.16 KHz, and $\omega_{high}$ = 70.64 KHz. At each drive frequency the free vibration amplitude $A_0$ is chosen such that multiple steady oscillation states are identifiable on the amplitude vs. distance and phase vs. distance curves. Interpretation of the amplitude and phase response of the tapping cantilever has been discussed extensively in the literature. Here we briefly discuss the observed force waveforms within the context of attractive and repulsive regimes, though, we are primarily interested in whether the steady state equations are quantitatively satisfied or not.

Figure (1) shows the amplitude and phase distance curves obtained at $\omega_{res}$ on a mica sample. At every data point on these curves, we obtained time-varying tip-sample forces and calculated corresponding phasors in Eq. (6). Due to the large number of data points we present our results (instantaneous amplitude, phase, tips-sample force waveform, and phasors) in movie format [27]. Here we have selected 4 representative data points on this curve that correspond to free vibration (A), attractive regime (B), hard tapping (C), and repulsive regime (D). For each selected data point the corresponding tip sample force waveform is also given in Fig. (1). In A, tip-sample force waveform is at the noise floor. In B, the forces are dominated by negative attractive forces, and in D repulsive forces dominate the interaction. The magnitude of the repulsive force is even larger in the hard tapping regime C. As expected, these waveforms qualitatively agree with the established understanding of the tapping-mode dynamics. We now show that time-varying forces quantitatively satisfy the steady state equation given in Eq. (6). For all data points on these curves we calculate the numerical values of the first and second terms of Eq. (6). The values of $F_{ts1}$ and $\theta$ are obtained from the corresponding tip-sample force waveform. Before plotting each term on the complex plane as phasors, we do normalization so that the magnitude of the driving force $F_D$ becomes unity. This helps to present the values in and around the unit circle. After normalization the magnitudes of each term are scaled as follows:

$$\begin{aligned} F_D &= K_1 A_0 T(\omega) \Rightarrow 1 \\ F_T &= K_1 A_s T(\omega) \Rightarrow A_s/A_0 \\ F_{ts1} &\Rightarrow F_{ts1}/F_D \end{aligned} \quad (7)$$

Fig. (1) shows the normalized phasor diagrams for the selected data points. Phasors calculated at all data points are presented in the supplementary movie 1 [28]. On a given phasor diagram the drive force **F_D**, tip-sample force **F_ts1**, and the summation **F_S** = **F_D** + **F_ts1** (left hand side of Eq. (1)) are depicted. If Eq. (6) is satisfied, the magnitude and phase of the summation **F_S** should match the magnitude $F_T$ and phase $\beta(\omega)$ of total force **F_T**. The normalized magnitude of $F_T$ (= $A_s/A_0$) is also depicted on each phasor diagram (black arrow on the horizontal axis) to aid this comparison.

On resonance phase $\beta(\omega_{res})$ is equal to 90 degrees. In the trivial free vibration case (A), $F_{ts1}$ is at the noise level and the drive force **F_D** is the total force. Note that the phase of **F_S** is approximately 90 degrees and its magnitude is equal to $F_T$ within the noise level (~ 1 nN RMS). In the case of attractive regime (B), **F_ts1** is aligned approximately with the horizontal axis with a phase close to 0 (in phase with the cantilever position). This is because attractive forces are negative in magnitude and they appear when the cantilever is at its lowest (most negative) point in its trajectory. Note that the phase of the drive force adjusts itself so that their sum **F<sub>S</sub>** remains at 90 degrees with a magnitude close to $F_T$ (compare red and black arrow lengths). In the hard tapping regime (C), **F<sub>ts1</sub>** is again close to the horizontal axis, but towards the negative direction. This is because the repulsive forces are positive in magnitude when the tip is at the bottom (out of phase). In this case **F<sub>D</sub>** has a phase less than 90 degrees. Again, the summation **F<sub>S</sub>** remains at 90 degrees with a magnitude close to $F_T$. In the repulsive regime (D) **F<sub>ts1</sub>** is still pointing towards the negative horizontal direction; however it has a smaller magnitude. Consequently **F<sub>D</sub>** adjusts its phase to a larger value and keeps their sum **F<sub>S</sub>** at 90 degrees.

It is important to observe the phasor diagram before and after inflection points where the cantilever oscillations switch between attractive and repulsive regimes (see movie 1 [28]). At those points the values of the right hand side terms of Eq (6) dramatically change while their sum remains the same.

We now discuss the measurements at $\omega_{high}$ = 75.12 KHz, and at $\omega_{low}$ = 74.07 KHz. Corresponding values of $\beta(\omega)$ and $T(\omega)$ are recalculated and the magnitudes of the phasors are normalized for each case. Figure 2 shows the data at $\omega_{high}$ in the same format as in Fig. 1. In this case $\beta(\omega)$ is larger than 90 degrees. Again, we selected four representative points on the curves for free vibration (A), attractive regime (B), hard tapping (C), and repulsive regime (D). In all four cases, **F<sub>S</sub>** is close to $F_T$ in magnitude and has a phase equal to that of the free vibration, i.e. $\beta(\omega_{high})$. Repulsive regime is less favorable at this drive frequency, consequently large **F<sub>ts1</sub>** values are observed (case C and D) as compared to on resonance case. Furthermore, stable tapping amplitudes larger than the free vibration amplitude are observed. Note that in this case **F<sub>T</sub>** has a larger magnitude than **F<sub>D</sub>**. Nevertheless, in all the cases the sum **F<sub>S</sub>** is close to **F<sub>T</sub>**. (See movie 2 for all the phasor diagrams at this frequency [28])

Finally, Fig. (3) shows the data obtained at $\omega_{low}$. The trivial free vibration case (A) shows that $\mathbf{F_S}$ coincides with $\mathbf{F_T}$ with a $\beta(\omega_{low})$ less than 90 degrees. In B, similar to the attractive regimes observed at $\omega_{res}$ and $\omega_{high}$ $\mathbf{F_{ts1}}$ is pointing towards the positive horizontal direction. The phase shift on the drive force is towards negative. Note that time-varying force waveforms (B) and (D) reveal interaction forces exhibiting repulsive components, while the overall interaction is attractive in nature. Attractive regimes involving intermittent contact were previously suggested [5]. Observing tip-sample force waveforms during imaging will allow to discriminate purely attractive regime from a dominant attractive regime; however it is beyond the current discussion. In all four selected cases here, the magnitude of $\mathbf{F_S}$ is close to $F_T$ and its phase remains around $\beta(\omega)$. (See movie 3 for all the measurements at this drive frequency [28])

In all three drive frequencies we measured that the absolute difference between the sum $\mathbf{F_S}$ and the total force $\mathbf{F_T}$ (right and left hand sides of Eq. (6)) has an RMS value les than 0.05 in the normalized scale calculated over all the data points. Considering the noise in the time-varying force measurements this implies that Eq. (6) is quantitatively satisfied to a good degree of accuracy. An important contributor to accuracy of these results is the calibration method of the torsional response of the THC presented in ref [27]. This method directly provides the ratio of the torsional and vertical response of the cantilever. The error associated with the calibration of the vertical spring constant is affecting all the terms in Eq. (6) in the same way and hence gets cancelled. Therefore, the accuracy of the results presented here are independent of the vertical spring constant calibration.

## IV. CONCLUSIONS

In this letter we presented experimental evidence that at multiple drive frequencies and tapping amplitudes independently measured time-varying tip-sample forces quantitatively satisfy the steady state equations for the cantilever dynamics. General characteristics of the measured tip-sample force waveforms in the attractive and repulsive regimes are also in agreement with the theoretical expectations. These results confirm the current understanding of the steady-state dynamics of the tapping-mode and build confidence on the reliability of time-varying tip-sample force measurements by the torsional harmonic cantilevers.

FIG. 1 (Color online) Results at $\omega_{res}$. (top row) tapping amplitude and phase vs. distance curves on mica. Force waveforms (A-D) and phasor diagrams (A-D) correspond to selected data points. Phasors: drive force $F_D$, tip-sample force $F_{ts1}$, and their sum $F_S$. The magnitude of the total force $F_T$ is given on the horizontal axis for comparison.

FIG. 2 (Color online) Results at $\omega_{high}$. (top row) tapping amplitude and phase vs. distance curves on mica. Force waveforms (A-D) and phasor diagrams (A-D) correspond to selected data points. Phasors: drive force $F_D$, tip-sample force $F_{ts1}$, and their sum $F_S$. The magnitude of the total force $F_T$ is given on the horizontal axis for comparison.

FIG. 3 (Color online) Results at $\omega_{low}$. (top row) tapping amplitude and phase vs. distance curves on mica. Force waveforms (A-D) and phasor diagrams (A-D) correspond to selected data points. Phasors: drive force $F_D$, tip-sample force $F_{ts1}$, and their sum $F_S$. The magnitude of the total force $F_T$ is given on the horizontal axis for comparison.

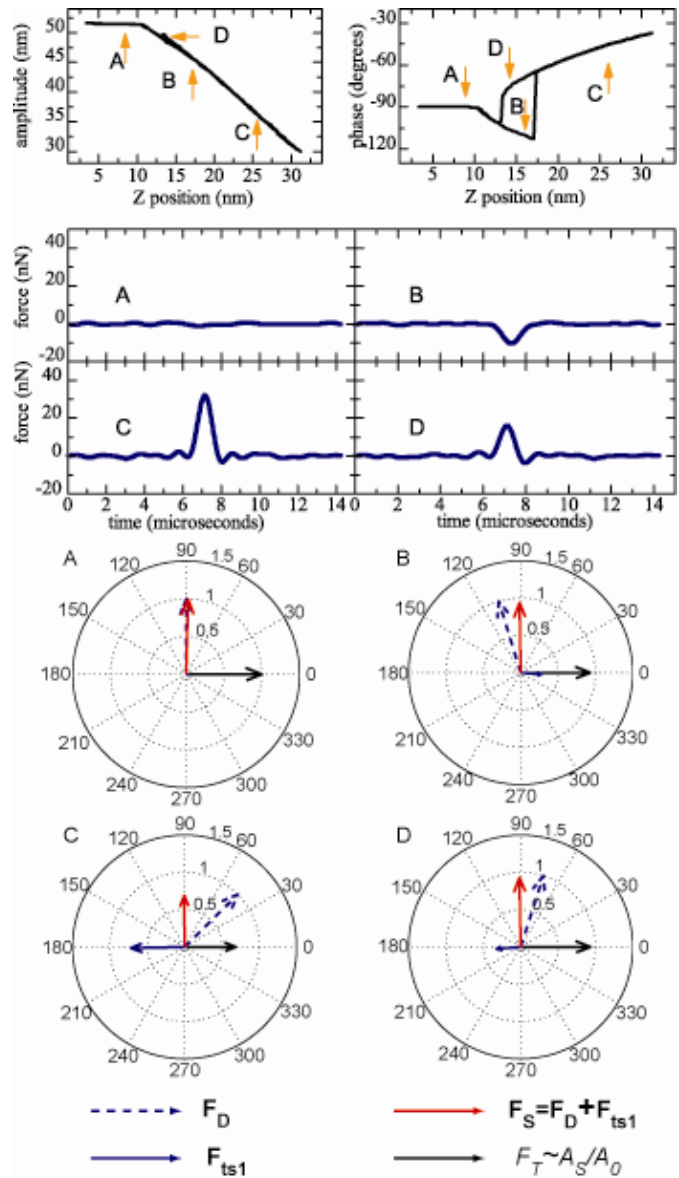

Fig. 1. O.Sahin

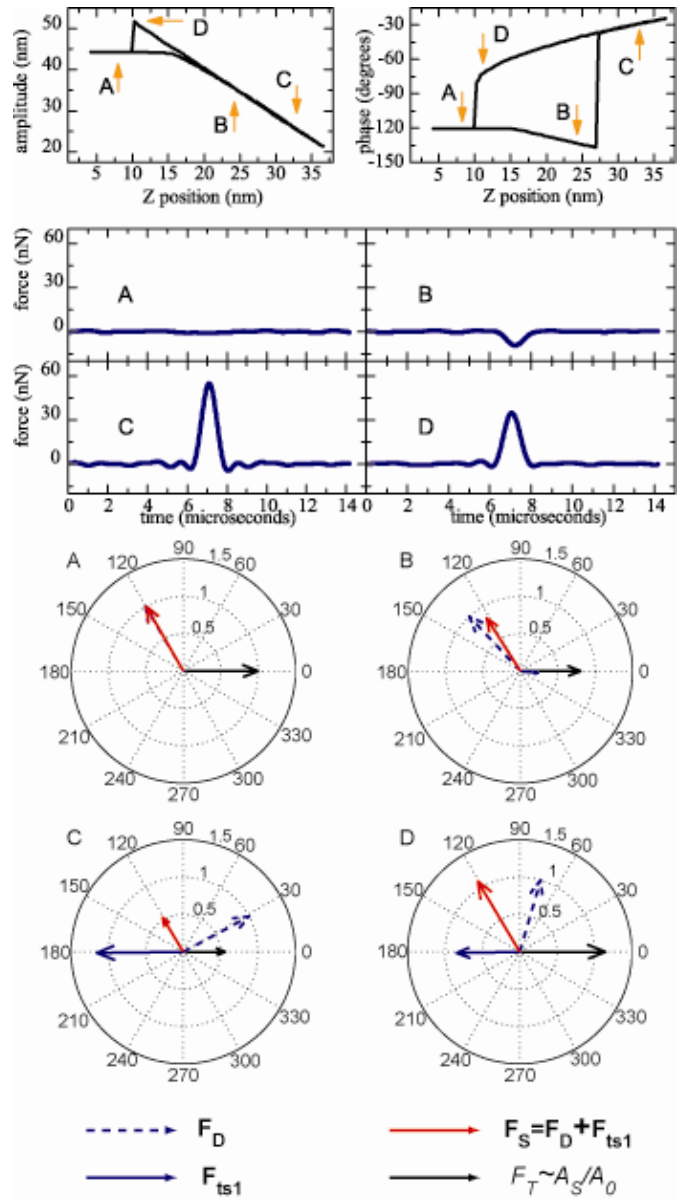

Fig. 2. O. Sahin

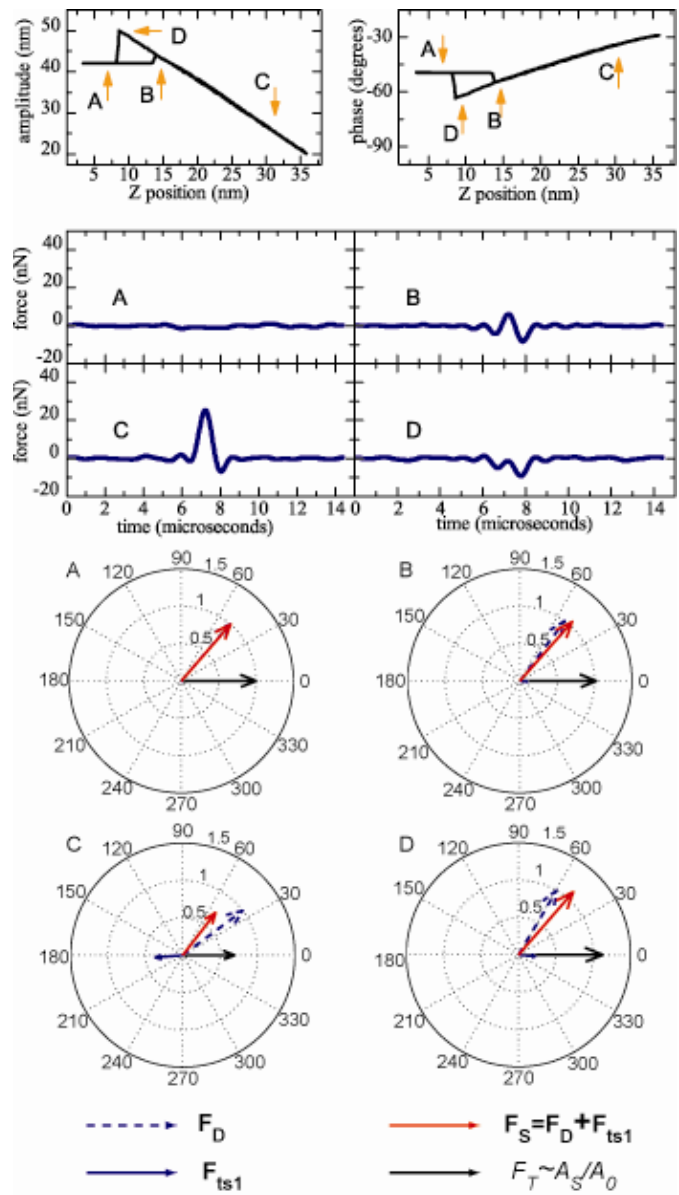

Fig. 3. O. Sahin